\documentclass[11pt]{amsart}
\usepackage{graphicx}
\usepackage{amscd}
\usepackage{amsbsy}
\usepackage{amsmath}
\usepackage{amstext}
\usepackage{amsfonts}
\usepackage{amssymb}
\newtheorem{theorem}{Theorem}
\theoremstyle{plain}

\newtheorem{definition}{Definition}

\newtheorem{proposition}{Proposition}
\newtheorem{remark}{Remark}

\numberwithin{equation}{section}

\makeindex

\begin{document}
\title[Grover's Algorithm]{A\\Lecture\\on\\Grover's Quantum Search Algorithm\\Version 1.1}
\author{Samuel J. Lomonaco, Jr.}
\address{Dept. of Comp. Sci. \& Elect. Engr.\\
University of Maryland Baltimore County\\
1000 Hilltop Circle\\
Baltimore, MD 21250}
\email{E-Mail: Lomonaco@UMBC.EDU}
\urladdr{WebPage: http://www.csee.umbc.edu/\symbol{126}lomonaco}
\thanks{This work was partially supported by ARO Grant \#P-38804-PH-QC and the L-O-O-P
Fund. The author gratefully acknowledges the hospitality of the University of
Cambridge Isaac Newton Institute for Mathematical Sciences, Cambridge,
England, where some of this work was completed. \ \ I would also like to thank
the other AMS\ Short Course lecturers, Howard Brandt, Dan Gottesman, Lou
Kauffman, Alexei Kitaev, Peter Shor, Umesh Vazirani and the many Short Course
participants for their support. \ (Copyright 2000.) \ }
\date{June 20, 2000}
\maketitle

\begin{abstract}
This paper ia a written version of a one hour lecture given on Lov Grover's
quantum database search algorithm. \ It is based on \cite{Grover1},
\cite{Grover2}, and \cite{Jozsa1}. \ 
\end{abstract}\tableofcontents

\section{Problem definition}

\qquad\bigskip

We consider the problem of searching an unstructured database of $N=2^{n}$
records for exactly one record which has been specifically marked. \ This can
be rephrased in mathematical terms as an oracle problem as follows:

\bigskip

Label the records of the database with the integers
\[
0,1,2,\ \ldots\ ,N-1\text{ ,}%
\]
and denote the label of the unknown marked record by $x_{0}$. \ We are given
\ an oracle which computes the $n$ bit binary function
\[
f:\left\{  0,1\right\}  ^{n}\longrightarrow\left\{  0,1\right\}
\]
defined by
\[
f(x)=\left\{
\begin{array}
[c]{cl}%
1 & \text{if }x=x_{0}\\
& \\
0 & \text{otherwise}%
\end{array}
\right.
\]

\bigskip

We remind the readers that, as a standard oracle idealization, we have no
access to the internal workings of the function $f$. \ It operates simply as a
blackbox function, which we can query as many times as we like. \ But with
each such a query comes an associated computational cost.

\bigskip

\noindent\textbf{Search Problem for an Unstructured Database.} \ \textit{Find
the record labeled as }$x_{0}$\textit{\ with the minimum amount of
computational work, i.e., with the minimum number of queries of the oracle
}$f$\textit{.}

\bigskip

From probability theory, we know that if we examine $k$ records, i.e., if we
compute the oracle $f$ for $k$ randomly chosen records, then the probability
of finding the record labeled as $x_{0}$ is $k/N$. \ Hence, on a classical
computer it takes $O(N)=O(2^{n})$ queries to find the record labeled $x_{0}$.

\bigskip

\section{The quantum mechanical perspective}

\qquad\bigskip

However, as Lov Grover so astutely observed, on a quantum computer the search
of an unstructured database can be accomplished in $O(\sqrt{N})$ steps, or
more precisely, with the application of $O(\sqrt{N}\lg N)$ sufficiently local
unitary transformations. \ Although this is not exponentially faster, it is a
significant speedup.

\vspace{0.5in}

Let $\mathcal{H}_{2}$ be a 2 dimensional Hilbert space with orthonormal basis
\[
\left\{  \left|  0\right\rangle ,\left|  1\right\rangle \right\}  \text{ ;}%
\]
and let
\[
\left\{  \left|  0\right\rangle ,\left|  1\right\rangle ,\ \ldots\ ,\left|
N-1\right\rangle \right\}
\]
denote the induced orthonormal basis of the Hilbert space
\[
\mathcal{H}=%
%TCIMACRO{\dbigotimes \limits_{0}^{N-1}}%
%BeginExpansion
{\displaystyle\bigotimes\limits_{0}^{N-1}}
%EndExpansion
\mathcal{H}_{2}\text{ .}%
\]

\vspace{0.5in}

From the quantum mechanical perspective, the oracle function $f$ is given as a
blackbox unitary transformation $U_{f}$, i.e., by
\[%
\begin{array}
[c]{ccc}%
\mathcal{H}\otimes\mathcal{H}_{2} & \overset{U_{f}}{\longrightarrow} &
\mathcal{H}\otimes\mathcal{H}_{2}\\
&  & \\
\left|  x\right\rangle \otimes\left|  y\right\rangle  & \longmapsto & \left|
x\right\rangle \otimes\left|  f(x)\oplus y\right\rangle
\end{array}
\]
where `$\oplus$' denotes exclusive `OR', i.e., addition modulo
2.\footnote{Please note that $U_{f}=\left(  \nu\circ\iota\right)  (f)$, as
defined in sections 10.3 and 10.4 of \cite{Lomonaco1}.}

\vspace{0.5in}

Instead of $U_{f}$, we will use the computationally equivalent unitary
transformation
\[
I_{\left|  x_{0}\right\rangle }\left(  \left|  x\right\rangle \right)
=(-1)^{f(x)}\left|  x\right\rangle =\left\{
\begin{array}
[c]{cl}%
-\left|  x_{0}\right\rangle  & \text{if \ }x=x_{0}\\
& \\
\left|  x\right\rangle  & \text{otherwise}%
\end{array}
\right.
\]
That $I_{\left|  x_{0}\right\rangle }$ is computationally equivalent to $U_{f}
$ follows from the easily verifiable fact that
\[
U_{f}\left(  \left|  x\right\rangle \otimes\frac{\left|  0\right\rangle
-\left|  1\right\rangle }{\sqrt{2}}\right)  =\left(  I_{\left|  x_{0}%
\right\rangle }\left(  \left|  x\right\rangle \right)  \right)  \otimes
\frac{\left|  0\right\rangle -\left|  1\right\rangle }{\sqrt{2}}\text{ ,}%
\]
and also from the fact that $U_{f}$ can be constructed from a
controlled-$I_{\left|  x_{0}\right\rangle }$ and two one qubit Hadamard
transforms. \ (For details, please refer to \cite{Jozsa3}, \cite{Kitaev1}.)

\vspace{0.5in}

The unitary transformation $I_{\left|  x_{0}\right\rangle }$ is actually an
\textbf{inversion} \cite{Beardon1} in $\mathcal{H}$ about the hyperplane
perpendicular to $\left|  x_{0}\right\rangle $. \ This becomes evident when
$I_{\left|  x_{0}\right\rangle }$ is rewritten in the form
\[
I_{\left|  x_{0}\right\rangle }=I-2\left|  x_{0}\right\rangle \left\langle
x_{0}\right|  \text{ ,}%
\]
where `$I$' denotes the identity transformation. \ More generally, for any
unit length ket $\left|  \psi\right\rangle $, the unitary transformation
\[
I_{\left|  \psi\right\rangle }=I-2\left|  \psi\right\rangle \left\langle
\psi\right|  \text{ }%
\]
is an inversion in $\mathcal{H}$ about the hyperplane orthogonal to $\left|
\psi\right\rangle $.

\vspace{0.5in}

\section{Properties of the inversion $I_{\left|  \psi\right\rangle }$}

\qquad\bigskip

We digress for a moment to discuss the properties of the unitary
transformation $I_{\left|  \psi\right\rangle }$. \ To do so, we need the
following definition.

\bigskip

\begin{definition}
Let $\left|  \psi\right\rangle $ and $\left|  \chi\right\rangle $ be two kets
in $\mathcal{H}$ for which the bracket product $\left\langle \psi\mid
\chi\right\rangle $ is a real number. \ We define
\[
\mathcal{S}_{\mathbb{C}}=Span_{\mathbb{C}}\left(  \left|  \psi\right\rangle
,\left|  \chi\right\rangle \right)  =\left\{  \alpha\left|  \psi\right\rangle
+\beta\left|  \chi\right\rangle \in\mathcal{H}\mid\alpha,\beta\in
\mathbb{C}\right\}
\]
as the sub-Hilbert space of $\mathcal{H}$ spanned by $\left|  \psi
\right\rangle $ and $\left|  \chi\right\rangle $. \ We associate with the
Hilbert space $\mathcal{S}_{\mathbb{C}}$ a real inner product space lying in
$\mathcal{S}_{\mathbb{C}}$ defined by
\[
\mathcal{S}_{\mathbb{R}}=Span_{\mathbb{R}}\left(  \left|  \psi\right\rangle
,\left|  \chi\right\rangle \right)  =\left\{  a\left|  \psi\right\rangle
+b\left|  \chi\right\rangle \in\mathcal{H}\mid a,b\in\mathbb{R}\right\}
\text{ ,}%
\]
where the inner product on $\mathcal{S}_{\mathbb{R}}$ is that induced by the
bracket product on $\mathcal{H}$. \ If $\left|  \psi\right\rangle $ and
$\left|  \chi\right\rangle $ are also linearly independent, then
$\mathcal{S}_{\mathbb{R}}$ is a 2 dimensional real inner product space (i.e.,
the 2 dimensional Euclidean plane) lying inside of the complex 2 dimensional
space $\mathcal{S}_{\mathbb{C}}$.
\end{definition}

\bigskip

\begin{proposition}
Let $\left|  \psi\right\rangle $ and $\left|  \chi\right\rangle $ be two
\ linearly independent unit length kets in $\mathcal{H}$ with real bracket
product; and let $\mathcal{S}_{\mathbb{C}}=Span_{\mathbb{C}}\left(  \left|
\psi\right\rangle ,\left|  \chi\right\rangle \right)  $ and $\mathcal{S}%
_{\mathbb{R}}=Span_{\mathbb{R}}\left(  \left|  \psi\right\rangle ,\left|
\chi\right\rangle \right)  $. \ Then

\begin{itemize}
\item [1)]Both $\mathcal{S}_{\mathbb{C}}$ and $\mathcal{S}_{\mathbb{R}}$ are
invariant under the transformations $I_{\left|  \psi\right\rangle }$,
$I_{\left|  \chi\right\rangle }$, and hence $I_{\left|  \psi\right\rangle
}\circ I_{\left|  \chi\right\rangle }$, i.e.,
\[
\fbox{$%
\begin{array}
[c]{lrl}%
I_{\left|  \psi\right\rangle }\left(  \mathcal{S}_{\mathbb{C}}\right)
=\mathcal{S}_{\mathbb{C}} & \ \text{and\ } & I_{\left|  \psi\right\rangle
}\left(  \mathcal{S}_{\mathbb{R}}\right)  =\mathcal{S}_{\mathbb{R}}\\
&  & \\
I_{\left|  \chi\right\rangle }\left(  \mathcal{S}_{\mathbb{C}}\right)
=\mathcal{S}_{\mathbb{C}} & \ \text{and\ } & I_{\left|  \chi\right\rangle
}\left(  \mathcal{S}_{\mathbb{R}}\right)  =\mathcal{S}_{\mathbb{R}}\\
&  & \\
I_{\left|  \psi\right\rangle }I_{\left|  \chi\right\rangle }\left(
\mathcal{S}_{\mathbb{C}}\right)  =\mathcal{S}_{\mathbb{C}} & \ \text{and\ } &
I_{\left|  \psi\right\rangle }I_{\left|  \chi\right\rangle }\left(
\mathcal{S}_{\mathbb{R}}\right)  =\mathcal{S}_{\mathbb{R}}%
\end{array}
$}%
\]
\bigskip

\item[2)] If $L_{\left|  \psi^{\perp}\right\rangle }$ is the line in the plane
$\mathcal{S}_{\mathbb{R}}$ which passes through the origin and which is
perpendicular to $\left|  \psi\right\rangle $, then $I_{\left|  \psi
\right\rangle }$ restricted to $\mathcal{S}_{\mathbb{R}}$ is a reflection in
(i.e., a M\"{o}bius inversion \cite{Beardon1} about) the line $L_{\left|
\psi^{\perp}\right\rangle }$. \ A similar statement can be made in regard to
$\left|  \chi\right\rangle $.

\item[3)] If $\left|  \psi^{\perp}\right\rangle $ is a unit length vector in
$\mathcal{S}_{\mathbb{R}}$ perpendicular to $\left|  \psi\right\rangle $,
then
\[
-I_{\left|  \psi\right\rangle }=I_{\left|  \psi^{\perp}\right\rangle }\text{
.}%
\]
(Hence, $\left\langle \psi^{\perp}\mid\chi\right\rangle $ is real.)
\end{itemize}
\end{proposition}

\vspace{0.5in}

Finally we note that, since $I_{\left|  \psi\right\rangle }=I-2\left|
\psi\right\rangle \left\langle \psi\right|  $, it follows that

\bigskip

\begin{proposition}
If $\left|  \psi\right\rangle $ is a unit length ket in $\mathcal{H}$, and if
$U$ is a unitary transformation on $\mathcal{H}$, then
\[
UI_{\left|  \psi\right\rangle }U^{-1}=I_{U\left|  \psi\right\rangle }\text{ .}%
\]
\end{proposition}

\vspace{0.5in}

\section{The method in Lov's ``madness''}

\qquad\bigskip

Let $H:\mathcal{H}\longrightarrow\mathcal{H}$ be the Hadamard transform,
i.e.,
\[
H=%
%TCIMACRO{\dbigotimes \limits_{0}^{n-1}}%
%BeginExpansion
{\displaystyle\bigotimes\limits_{0}^{n-1}}
%EndExpansion
H^{(2)}\text{ , }%
\]
where
\[
H^{(2)}=\left(
\begin{array}
[c]{rr}%
1 & 1\\
1 & -1
\end{array}
\right)
\]
with respect to the basis $\left|  0\right\rangle $, $\left|  1\right\rangle $.

\bigskip

We begin by using the Hadamard transform $H$ to construct a state $\left|
\psi_{0}\right\rangle $ which is an equal superposition of all the standard
basis states $\left|  0\right\rangle $, $\left|  1\right\rangle $,$\ldots
$,$\left|  N-1\right\rangle $ (including the unknown state $\left|
x_{0}\right\rangle $), i.e.,
\[
\left|  \psi_{0}\right\rangle =H\left|  0\right\rangle =\frac{1}{\sqrt{N}}%
\sum_{k=0}^{N-1}\left|  k\right\rangle \text{ .}%
\]

\bigskip

Both $\left|  \psi_{0}\right\rangle $ and the unknown state $\left|
x_{0}\right\rangle $ lie in the Euclidean plane $\mathcal{S}_{\mathbb{R}%
}=Span_{\mathbb{R}}\left(  \left|  \psi_{0}\right\rangle ,\left|
x_{0}\right\rangle \right)  $. \ Our strategy is to rotate within the plane
$\mathcal{S}_{\mathbb{R}}$ the state $\left|  \psi_{0}\right\rangle $ about
the origin until it is as close as possible to $\left|  x_{0}\right\rangle $.
\ Then a measurement with respect to the standard basis of the state resulting
from rotating $\left|  \psi_{0}\right\rangle $, will produce $\left|
x_{0}\right\rangle $ with high probability.

\bigskip

To achieve this objective, we use the oracle $I_{\left|  x_{0}\right\rangle }$
to construct the unitary transformation
\[
Q=-HI_{\left|  0\right\rangle }H^{-1}I_{\left|  x_{0}\right\rangle }\text{ ,}%
\]

\bigskip

which by proposition 2 above, can be reexpressed as
\[
Q=-I_{\left|  \psi_{0}\right\rangle }I_{\left|  x_{0}\right\rangle }\text{ .}%
\]

\bigskip

Let $\left|  x_{0}^{\perp}\right\rangle $ and $\left|  \psi_{0}^{\perp
}\right\rangle $ denote unit length vectors in $\mathcal{S}_{\mathbb{R}}$
perpendicular to $\left|  x_{0}\right\rangle $ and $\left|  \psi
_{0}\right\rangle $, respectively. \ There are two possible choices for each
of $\left|  x_{0}^{\perp}\right\rangle $ and $\left|  \psi_{0}^{\perp
}\right\rangle $ respectively. \ To remove this minor, but nonetheless
annoying, ambiguity, we select $\left|  x_{0}^{\perp}\right\rangle $ and
$\left|  \psi_{0}^{\perp}\right\rangle $ so that the orientation of the plane
$\mathcal{S}_{\mathbb{R}}$ induced by the ordered spanning vectors $\left|
\psi_{0}\right\rangle $, $\left|  x_{0}\right\rangle $ is the same orientation
as that induced by each of the ordered bases $\left|  x_{0}^{\perp
}\right\rangle $, $\left|  x_{0}\right\rangle $ and $\left|  \psi
_{0}\right\rangle $, $\left|  \psi_{0}^{\perp}\right\rangle $. \ (Please refer
to Figure 2.)

\bigskip

\begin{remark}
The removal of the above ambiguities is really not essential. \ However, it
does simplify the exposition given below.
\end{remark}

\bigskip%

%TCIMACRO{\FRAME{dtbpFUX}{2.8323in}{1.8248in}{0pt}{\Qcb{Figure 2. \ The linear
%transformation $\left.  Q\right|  _{\mathcal{S}_{\mathbb{R}}}$ is reflection
%in the line $L_{\left|  x_{0}^{\perp}\right\rangle }$ followed by reflection
%in the line $L_{\left|  \psi_{0}\right\rangle }$ which is the same as rotation
%by the angle $2\beta$. \ Thus, $\left.  Q\right|  _{\mathcal{S}_{\mathbb{R}}}$
%rotates $\left|  \psi_{0}\right\rangle $ by the angle $2\beta$ toward $\left|
%x_{0}\right\rangle $. }}{}{mobius.ps}{\special{ language "Scientific Word";
%type "GRAPHIC";  maintain-aspect-ratio TRUE;  display "USEDEF";
%valid_file "F";  width 2.8323in;  height 1.8248in;  depth 0pt;
%original-width 7.1131in;  original-height 4.574in;  cropleft "0";
%filename 'mobius.ps';file-properties "XNPEU";}}}%
%BeginExpansion
\begin{center}
\fbox{\includegraphics[
height=1.8248in,
width=2.8323in
]%
{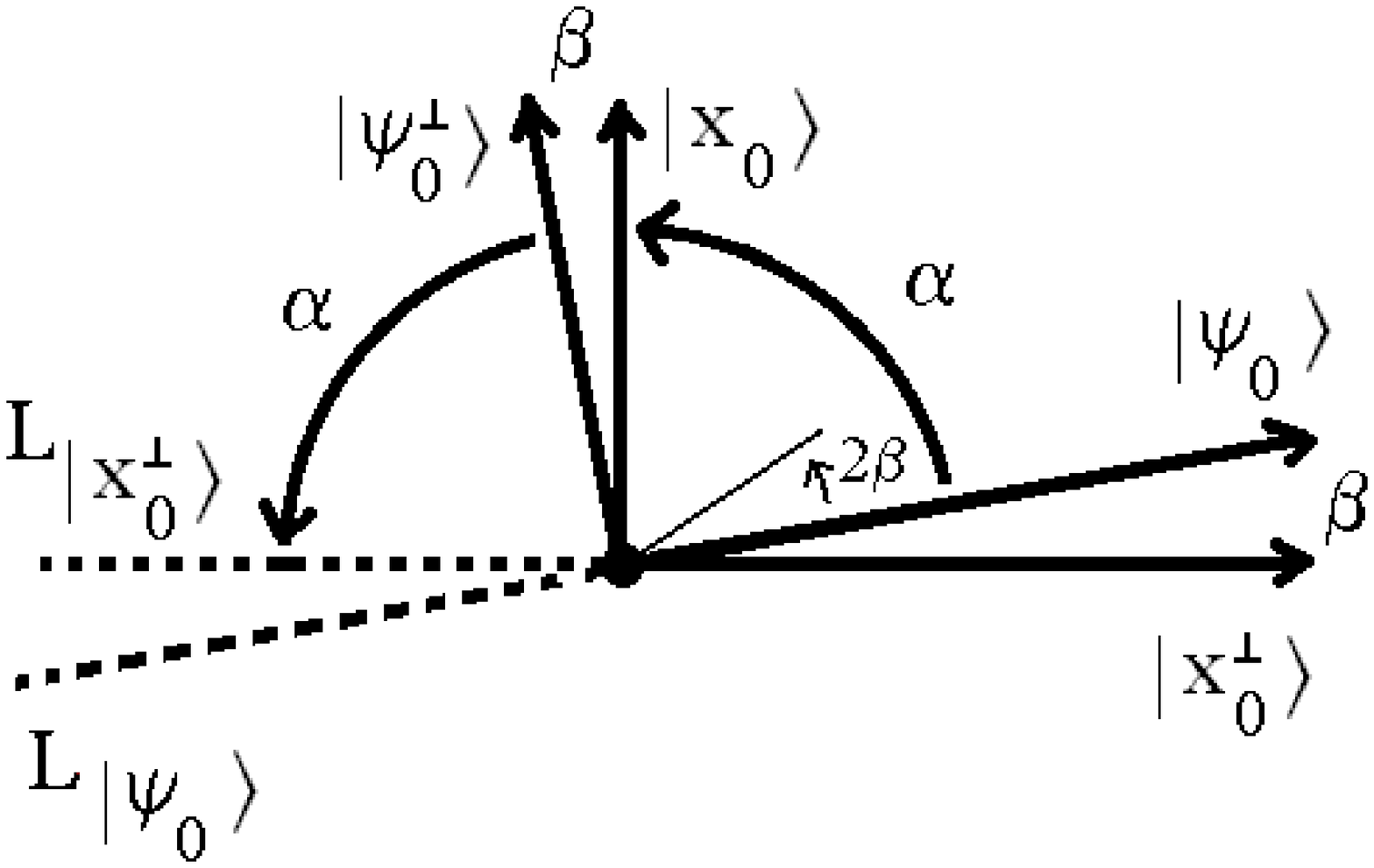}%
}\\
Figure 2. \ The linear transformation $\left.  Q\right|  _{\mathcal{S}%
_{\mathbb{R}}}$ is reflection in the line $L_{\left|  x_{0}^{\perp
}\right\rangle }$ followed by reflection in the line $L_{\left|  \psi
_{0}\right\rangle }$ which is the same as rotation by the angle $2\beta$.
\ Thus, $\left.  Q\right|  _{\mathcal{S}_{\mathbb{R}}}$ rotates $\left|
\psi_{0}\right\rangle $ by the angle $2\beta$ toward $\left|  x_{0}%
\right\rangle $.
\end{center}
%EndExpansion

\vspace{0.5in}

We proceed by noting that, by the above proposition 1, the plane
$\mathcal{S}_{\mathbb{R}}$ lying in $\mathcal{H}$ is invariant under the
linear transformation $Q$, and that, when $Q$ is restricted to the plane
$\mathcal{S}_{\mathbb{R}}$, it can be written as the composition of two
inversions, i.e.,
\[
\left.  Q\right|  _{\mathcal{S}_{\mathbb{R}}}=I_{\left|  \psi_{0}^{\perp
}\right\rangle }I_{\left|  x_{0}\right\rangle }\text{ .}%
\]

In particular, $\left.  Q\right|  _{\mathcal{S}_{\mathbb{R}}}$ is the
composition of two inversions in $\mathcal{S}_{\mathbb{R}}$, the first in the
line $L_{\left|  x_{0}^{\perp}\right\rangle }$ in $\mathcal{S}_{\mathbb{R}}$
passing through the origin having $\left|  x_{0}\right\rangle $ as normal, the
second in the line $L_{\left|  \psi_{0}\right\rangle }$ through the origin
having $\left|  \psi_{0}^{\perp}\right\rangle $ as normal.\footnote{The line
$L_{\left|  x_{0}^{\perp}\right\rangle }$ is the intersection of the plane
$\mathcal{S}_{\mathbb{R}}$ with the hyperplane in $\mathcal{H}$ orthogonal to
$\left|  x_{0}\right\rangle $. \ A similar statement can be made in regard to
$L_{\left|  \psi_{0}\right\rangle }$.} \ 

\vspace{0.5in}

We can now apply the following theorem from plane geometry:

\bigskip

\begin{theorem}
If $L_{1}$ and $L_{2}$ are lines in the Euclidean plane $\mathbb{R}^{2}$
intersecting at a point $O$; and if $\beta$ is the angle in the plane from
$L_{1}$ to $L_{2}$, then the operation of reflection in $L_{1}$ followed by
reflection in $L_{2}$ is just rotation by angle $2\beta$ about the point $O$.
\end{theorem}

\bigskip

Let $\beta$ denote the angle in $S_{\mathbb{R}}$ from $L_{\left|  x_{0}%
^{\perp}\right\rangle }$ to $L_{\left|  \psi_{0}\right\rangle }$, which by
plane geometry is the same as the angle from $\left|  x_{0}^{\perp
}\right\rangle $ to $\left|  \psi_{0}\right\rangle $, which in turn is the
same as the angle from $\left|  x_{0}\right\rangle $ to $\left|  \psi
_{0}^{\perp}\right\rangle $. \ Then by the above theorem $\left.  Q\right|
_{\mathcal{S}_{\mathbb{R}}}=I_{\left|  \psi_{0}^{\perp}\right\rangle
}I_{\left|  x_{0}\right\rangle }$ is a rotation about the origin by the angle
$2\beta$. \ 

\vspace{0.5in}

The key idea in Grover's algorithm is to move $\left|  \psi_{0}\right\rangle $
toward the unknown state $\left|  x_{0}\right\rangle $ by successively
applying the rotation $Q$ to $\left|  \psi_{0}\right\rangle $ to rotate it
around to $\left|  x_{0}\right\rangle $. \ This process is called
\textbf{amplitude amplification}. \ \ Once this process is completed, the
measurement of the resulting state (with respect to the standard basis) will,
with high probability, yield the unknown state $\left|  x_{0}\right\rangle $.
\ This is the essence of Grover's algorithm. \ 

\vspace{0.5in}

But how many times $K$ should we apply the rotation $Q$ to $\left|  \psi
_{0}\right\rangle $? \ If we applied $Q$ too many or too few times, we would
over- or undershoot our target state $\left|  x_{0}\right\rangle $. \ 

\vspace{0.5in}

We determine the integer $K$ as follows:

\bigskip

Since
\[
\left|  \psi_{0}\right\rangle =\sin\beta\left|  x_{0}\right\rangle +\cos
\beta\left|  x_{0}^{\perp}\right\rangle \text{ ,}%
\]
the state resulting after $k$ applications of $Q$ is
\[
\left|  \psi_{k}\right\rangle =Q^{k}\left|  \psi_{0}\right\rangle =\sin\left[
\left(  2k+1\right)  \beta\right]  \left|  x_{0}\right\rangle +\cos\left[
\left(  2k+1\right)  \beta\right]  \left|  x_{0}^{\perp}\right\rangle \text{
.}%
\]
Thus, we seek to find the smallest positive integer $K=k$ such that
\[
\sin\left[  \left(  2k+1\right)  \beta\right]
\]
is as close as possible to $1$. \ In other words, we seek to find the smallest
positive integer $K=k$ such that
\[
\left(  2k+1\right)  \beta
\]
is as close as possible to $\pi/2$. \ It follows that\footnote{The reader may
prefer to use the $floor$ function instead of the $round$ function.}
\[
K=k=round\left(  \frac{\pi}{4\beta}-\frac{1}{2}\right)  \text{ ,}%
\]
where ``$round$'' is the function that rounds to the nearest integer.

\vspace{0.5in}

We can determine the angle $\beta$ by noting that the angle $\alpha$ from
$\left|  \psi_{0}\right\rangle $ and $\left|  x_{0}\right\rangle $ is
complementary to $\beta$, i.e.,
\[
\alpha+\beta=\pi/2\text{ ,}%
\]
and hence,
\[
\frac{1}{\sqrt{N}}=\left\langle x_{0}\mid\psi_{0}\right\rangle =\cos
\alpha=\cos(\frac{\pi}{2}-\beta)=\sin\beta\text{ .}%
\]
Thus, the angle $\beta$ is given by
\[
\beta=\sin^{-1}\left(  \frac{1}{\sqrt{N}}\right)  \approx\frac{1}{\sqrt{N}%
}\text{ \ (for large }N\text{) ,}%
\]
and hence,
\[
K=k=round\left(  \frac{\pi}{4\sin^{-1}\left(  \frac{1}{\sqrt{N}}\right)
}-\frac{1}{2}\right)  \approx round\left(  \frac{\pi}{4}\sqrt{N}-\frac{1}%
{2}\right)  \text{ (for large }N\text{).}%
\]

\vspace{0.5in}

\section{Summary of Grover's algorithm}

\qquad\bigskip

In summary, we provide the following outline of Grover's algorithm:

\bigskip

\fbox{%
\begin{tabular}
[c]{ll}\hline\hline
& \hspace{0.75in}\textbf{Grover's Algorithm}\\\hline\hline
& \\
$%
\begin{array}
[c]{r}%
\fbox{$\mathbb{STEP}$ 0.}\\
\bigskip\\
\bigskip
\end{array}
$ & $%
\begin{array}
[c]{l}%
\text{(Initialization)}\\
\qquad\left|  \psi\right\rangle \longleftarrow H\left|  0\right\rangle
=\frac{1}{\sqrt{N}}%
%TCIMACRO{\dsum \limits_{j=0}^{N-1}}%
%BeginExpansion
{\displaystyle\sum\limits_{j=0}^{N-1}}
%EndExpansion
\left|  j\right\rangle \\
\qquad k\quad\longleftarrow0
\end{array}
$\\
& \\
$%
\begin{array}
[c]{r}%
\fbox{$\mathbb{STEP}$ 1.}\\
\bigskip\\
\bigskip
\end{array}
$ & $%
\begin{array}
[c]{r}%
\text{Loop until }k=\underset{}{round\left(  \frac{\pi}{4\sin^{-1}\left(
1/\sqrt{N}\right)  }-\frac{1}{2}\right)  }\approx round\left(  \frac{\pi}%
{4}\sqrt{N}-\frac{1}{2}\right) \\
\multicolumn{1}{l}{\qquad\left|  \psi\right\rangle \longleftarrow\underset
{}{Q}\left|  \psi\right\rangle =-HI_{\left|  0\right\rangle }HI_{\left|
x_{0}\right\rangle }\left|  \psi\right\rangle }\\
\multicolumn{1}{l}{\qquad k\quad\longleftarrow k+1}%
\end{array}
$\\
& \\
$%
\begin{array}
[c]{r}%
\fbox{$\mathbb{STEP}$ 2.}\\
\bigskip
\end{array}
$ & $%
\begin{array}
[c]{l}%
\text{Measure }\left|  \psi\right\rangle \text{ with respect to the standard
basis}\\
\left|  0\right\rangle ,\left|  1\right\rangle ,\ \ldots\ ,\left|
N-1\right\rangle \text{ to obtain the marked unknown }\\
\text{state }\left|  x_{0}\right\rangle \text{ with probability }\geq
1-\frac{1}{N}\text{.}%
\end{array}
$%
\end{tabular}
}

\vspace{0.5in}

We complete our summary with the following theorem:

\bigskip

\begin{theorem}
With a probability of error\footnote{If the reader prefers to use the $floor$
function rather than the $round$ function, then probability of error becomes
$Prob_{E}\leq\frac{4}{N}-\frac{4}{N^{2}}$.}
\[
Prob_{E}\leq\frac{1}{N}\text{, }%
\]
Grover's algorithm finds the unknown state $\left|  x_{0}\right\rangle $ at a
computational cost of
\[
O\left(  \sqrt{N}\lg N\right)
\]
\end{theorem}

\begin{proof}
\qquad\bigskip

\begin{itemize}
\item [Part 1.]The probability of error $Prob_{E}$ of finding the hidden state
$\left|  x_{0}\right\rangle $ is given by
\[
Prob_{E}=\cos^{2}\left[  \left(  2K+1\right)  \beta\right]  \text{ ,}%
\]
where
\[
\left\{
\begin{array}
[c]{rrl}%
\beta & = & \sin^{-1}\left(  \frac{1}{\sqrt{N}}\right) \\
&  & \\
K & = & round\left(  \frac{\pi}{4\beta}-\frac{1}{2}\right)
\end{array}
\right.  \text{,}%
\]
where ``$round$'' is the function that rounds to the nearest integer. Hence,
\[%
\begin{array}
[c]{rrl}%
\frac{\pi}{4\beta}-1\leq K\leq\frac{\pi}{4\beta} & \Longrightarrow & \frac
{\pi}{2}-\beta\leq\left(  2K+1\right)  \beta\leq\frac{\pi}{2}+\beta\\
&  & \\
& \Longrightarrow & \sin\beta=\cos\left(  \frac{\pi}{2}-\beta\right)  \geq
\cos\left[  \left(  2K+1\right)  \beta\right]  \geq\cos\left(  \frac{\pi}%
{2}+\beta\right)  =-\sin\beta
\end{array}
\]
Thus,
\[
Prob_{E}=\cos^{2}\left[  \left(  2K+1\right)  \beta\right]  \leq\sin^{2}%
\beta=\sin^{2}\left(  \sin^{-1}\left(  \frac{1}{\sqrt{N}}\right)  \right)
=\frac{1}{N}%
\]
\end{itemize}

\bigskip

\begin{itemize}
\item [Part 2.]The computational cost of the Hadamard transform $H=\bigotimes
_{0}^{n-1}H^{(2)}$ is $O(n)=O(\lg N)$ single qubit operations. \ The
transformations $-I_{\left|  0\right\rangle }$ and $I_{\left|  x_{0}%
\right\rangle }$ each carry a computational cost of $O(1)$.

$\mathbb{STEP}$ 1 is the computationally dominant step. \ In $\mathbb{STEP}$ 1
there are $O\left(  \sqrt{N}\right)  $ iterations. \ In each iteration, the
Hadamard transform is applied twice. \ The transformations $-I_{\left|
0\right\rangle }$ and $I_{\left|  x_{0}\right\rangle }$ are each applied once.
Hence, each iteration comes with a computational cost of $O\left(  \lg
N\right)  $, and so the total cost of $\mathbb{STEP}$ 1 is $O(\sqrt{N}\lg N)$.
\end{itemize}
\end{proof}

\bigskip

\section{\textbf{\bigskip}An example of Grover's algorithm}

\qquad\bigskip

As an example, we search a database consisting of $N=2^{n}=8$ records for an
unknown record with the unknown label $x_{0}=5$. \ The calculations for this
example were made with OpenQuacks, which is an open source quantum simulator
Maple package developed at UMBC and publically available.

\vspace{0.5in}

We are given a blackbox computing device
\[
\text{In}\rightarrow\fbox{\fbox{%
\begin{tabular}
[c]{l}%
$I_{\left|  ?\right\rangle }$%
\end{tabular}
}}\rightarrow\text{Out}%
\]
that implements as an oracle the unknown unitary transformation
\[
I_{\left|  x_{0}\right\rangle }=I_{\left|  5\right\rangle }=\left(
\begin{array}
[c]{rrrrrrrrr}%
1 & 0 & 0 & 0 &  & 0 & 0 & 0 & 0\\
0 & 1 & 0 & 0 &  & 0 & 0 & 0 & 0\\
0 & 0 & 1 & 0 &  & 0 & 0 & 0 & 0\\
0 & 0 & 0 & 1 &  & 0 & 0 & 0 & 0\\
&  &  &  &  &  &  &  & \\
0 & 0 & 0 & 0 &  & -1 & 0 & 0 & 0\\
0 & 0 & 0 & 0 &  & 0 & 1 & 0 & 0\\
0 & 0 & 0 & 0 &  & 0 & 0 & 1 & 0\\
0 & 0 & 0 & 0 &  & 0 & 0 & 0 & 1
\end{array}
\right)
\]

\bigskip

We cannot open up the blackbox $\rightarrow\fbox{$\fbox{%
\begin{tabular}
[c]{l}%
$I_{\left|  ?\right\rangle }$%
\end{tabular}
}$}\rightarrow$ to see what is inside. \ So we do not know what $I_{\left|
x_{0}\right\rangle }$ and $x_{0}$ are. \ \ The only way that we can glean some
information about $x_{0}$ is to apply some chosen state $\left|
\psi\right\rangle $ as input, and then make use of the resulting output.

\vspace{0.5in}

Using of the blackbox $\rightarrow\fbox{\fbox{%
\begin{tabular}
[c]{l}%
$I_{\left|  ?\right\rangle }$%
\end{tabular}
}}\rightarrow$ as a component device, we construct a computing device
$\rightarrow\fbox{\fbox{%
\begin{tabular}
[c]{l}%
$-HI_{\left|  0\right\rangle }HI_{\left|  ?\right\rangle }$%
\end{tabular}
}}\rightarrow$ which implements the unitary operator
\[
Q=-HI_{\left|  0\right\rangle }HI_{\left|  x_{0}\right\rangle }=\frac{1}%
{4}\left(
\begin{array}
[c]{rrrrrrrrr}%
-3 & 1 & 1 & 1 &  & -1 & 1 & 1 & 1\\
1 & -3 & 1 & 1 &  & -1 & 1 & 1 & 1\\
1 & 1 & -3 & 1 &  & -1 & 1 & 1 & 1\\
1 & 1 & 1 & -3 &  & -1 & 1 & 1 & 1\\
&  &  &  &  &  &  &  & \\
1 & 1 & 1 & 1 &  & 3 & 1 & 1 & 1\\
1 & 1 & 1 & 1 &  & -1 & -3 & 1 & 1\\
1 & 1 & 1 & 1 &  & -1 & 1 & -3 & 1\\
1 & 1 & 1 & 1 &  & -1 & 1 & 1 & -3
\end{array}
\right)
\]

\vspace{0.5in}

We do not know what unitary transformation $Q$ \ is implemented by the device
$\rightarrow\fbox{\fbox{%
\begin{tabular}
[c]{l}%
$-HI_{\left|  0\right\rangle }HI_{\left|  ?\right\rangle }$%
\end{tabular}
}}\rightarrow$ because the blackbox $\rightarrow\fbox{\fbox{%
\begin{tabular}
[c]{l}%
$I_{\left|  ?\right\rangle }$%
\end{tabular}
}}\rightarrow$ is one of its essential components.

\bigskip

\begin{itemize}
\item [\fbox{$\mathbb{STEP}$ 0.}]We begin by preparing the known state
\[
\fbox{$\left|  \psi_0\right\rangle =H\left|  0\right\rangle =\frac{1}{\sqrt
{8}}\left(  1,1,1,1,1,1,1,1\right)  ^{transpose}$}%
\]
\end{itemize}

\bigskip

\begin{itemize}
\item [\fbox{$\mathbb{STEP}$ 1.}]We proceed to loop
\[
K=round\left(  \frac{\pi}{4\sin^{-1}\left(  1/\sqrt{8}\right)  }-\frac{1}%
{2}\right)  =2
\]
times in $\mathbb{STEP}$ 1.

\begin{itemize}
\item [\textsc{Iteration} 1.]On the first iteration, we obtain the unknown
state
\[
\fbox{$\left|  \psi_1\right\rangle =Q\left|  \psi_0\right\rangle =\frac
{1}{4\sqrt{2}}\left(  1,1,1,1,5,1,1,1\right)  ^{transpose}$}%
\]

\item[\textsc{Iteration} 2.] On the second iteration, we obtain the unknown
state
\[
\fbox{$\left|  \psi_2\right\rangle =Q\left|  \psi_1\right\rangle =\frac
{1}{8\sqrt{2}}\left(  -1,-1,-1,-1,11,-1,-1,-1\right)  ^{transpose}$}%
\]
and branch to $\mathbb{STEP}$ 2.
\end{itemize}
\end{itemize}

\bigskip

\begin{itemize}
\item [\fbox{$\mathbb{STEP}$ 2.}]We measure the unknown state $\left|
\psi_{2}\right\rangle $ to obtain either
\[
\left|  5\right\rangle
\]
with probability
\[
Prob_{Success}=\sin^{2}\left(  \left(  2K+1\right)  \beta\right)  =\frac
{121}{128}=0.9453
\]
or some other state with probability
\[
Prob_{Failure}=\cos^{2}\left(  \left(  2K+1\right)  \beta\right)  =\frac
{7}{128}=0.0547
\]
and then exit.
\end{itemize}

\quad
\end{document}